\newcommand*{\BODKIM}{}%

\ifdefined\BODKIM
\documentclass[journal,draftcls,onecolumn,12pt,twoside]{IEEEtranTCOM}
\else
\documentclass[journal]{IEEEtran}
\fi

\ifdefined\BODKIM
\newcommand\figSize{1}
\else
\newcommand\figSize{0.85}
\fi

\usepackage{cite}
\usepackage{graphicx}
\usepackage{amssymb,amsmath}
\usepackage{amsfonts}
\usepackage{float} % is used for \begin{figure}{H}
\usepackage{srcltx}
\usepackage{mathrsfs} % is required for \mathscr alphabet.
\usepackage{multirow}
\usepackage{bbm}
\usepackage{cuted}
\usepackage{amsthm}

\usepackage{graphicx}
\usepackage{epsfig}
\usepackage{psfrag}

\usepackage{setspace}
\usepackage{makecell}

\usepackage[]{units} % for nicefrac
\usepackage{url} % for URL bib
\usepackage[dvips]{color}
\usepackage{verbatim} % for comment environment

\newcommand{\svv}[1]{\mathbf{#1}}
\DeclareRobustCommand{\Nt}[1][Nt]{\ensuremath {N_t}}
\DeclareRobustCommand{\alNt}[1][Nt]{\alpha(\Nt)}
\DeclareRobustCommand{\aNorm}[1][aNorm]{\ensuremath {\|{\bf a}\|}}
\DeclareRobustCommand{\PrecMat}[1][Nt]{\ensuremath{G}}

\newcommand{\SNR}{\text{$\mathsf{SNR}$}}

\DeclareRobustCommand{\prob}[1][{\rm Pr}]{\ensuremath {{#1}}}

\begin{document}

\allowdisplaybreaks

\title{A Simple Receive Diversity Technique for Distributed Beamforming}
% \author{
% \IEEEauthorblockN{Elad Domanovitz and Uri Erez}
% \\
% \IEEEauthorblockA{%Dept. of EE-Systems,
% %TAU\\
% Dept. EE-Systems, Tel Aviv University, Israel
% }
% }
%\author{Author 1 and Author 2
\author{Elad Domanovitz and Uri Erez

\thanks{The work of E. Domanovitz and U. Erez was supported in part by the Israel Science Foundation under Grant No. 1956/17.}
% %\thanks{The material in this paper was presented in part at the 2016 IEEE
% %International Symposium on Information Theory, Barcelona.}
\thanks{E. Domanovitz and U. Erez are with the Department of Electrical Engineering -- Systems, Tel Aviv University, Tel Aviv, Israel (email: domanovi,uri@eng.tau.ac.il).}
}
\maketitle
% \date{September 2017}

\begin{abstract}
A simple generalization of distributed beamforming is proposed for use in a scenario involving a single-antenna source node communicating with a destination node that is equipped with two antennas via multiple relay nodes, each of which may be equipped with an arbitrary number of antennas.
%is subject to an individual power constraint. Furthermore, ultra-reliable and low-latency communication are desired. The latter requirement translates to considering only schemes that make use of local channel state information. Whereas for a receiver equipped with a single antenna, distributed beamforming is a well known and adequate solution, no straightforward extension is known.
The  proposed method is based on  a space-time diversity transformation that is applied as a front-end operation at the destination node, resulting in an effective unitary channel matrix between each relay and the destination. 
%replacing the scalar  coefficient corresponding to each user. 
Each relay node then inverts its
associated channel matrix and then forwards the decoded codeword, or received signal, over the resulting ``gain-only" channel. 
%The proposed scheme may be viewed as the natural generalization of distributed beamforming. 
%, and then repeats the message over the resulting ``gain-only" channel. 
%In comparison to a single-antenna destination node, the method doubles the diversity order without requiring any channel state information at the receiver while at the same time retaining the array gain offered by the relays. 
\end{abstract}

\section{Introduction}
Cooperative diversity is a means to boost the reliability of communication over a wireless channel where adjacent devices collaborate and share their antennas to facilitate communication between a source and destination node.
Different approaches and transmission protocols have been investigated over the years to address this goal.

The potential of using multiple single-antenna relay nodes as a means of forming a virtual antenna array has been recognized and studied in depth
since the pioneering work of
\cite{sendonaris2003user1,laneman2003distributed}. Depending on the assumptions made on the availability of channel state information (CSI) at the relays, the virtual antenna array can serve either to provide diversity alone or to obtain also array (power) gain. The former goal does not require (forward channel, from relay to destination) CSI at the relays and may be achieved via 
distributed space-time coding as suggested in \cite{laneman2003distributed}, or by means of
opportunistic relay selection as suggested in
\cite{bletsas2006simple}.
%In general, achieving the latter goal (array gain) requires full CSI to be available at the relays. This requirement precludes using these methods in case low latency is required as achieving full CSI is a time consuming process.
In contrast, achieving the potential array gain offered by the relays requires at least local CSI to be available to the relays.

It is well known that in the case of a system where all nodes are equipped with a single antenna and each relay knows the channel gain between itself and the destination, distributed (phase-only) beamforming  offers both diversity and array gain \cite{sendonaris2003user1}. In fact, only a small power loss, with respect to the full (centralized) array gain, is incurred  by the availability of only local CSI. Specifically, the loss is identical to that incurred in the dual (receiver side) scenario of performing equal-gain combining in place of maximal-ratio combining (MRC).
%which is a classic problem that has been explored in depth \cite{brennan1959linear}. 
Further, it has been shown in \cite{zheng2007mimo} that given  a per-relay power constraint, such phase-only beamforming is optimal (in the sense of maximizing the receive SNR).
In the present paper, we extend distributed beamforming
%the insight of \cite{sendonaris2003user1} (i.e., that phase-only beamforming loses little with respect to centralized beamforming) 
to a scenario  where the destination node is equipped with two receive antennas.

Recently,
%as interest in Internet-of-Things %(IoT) scenarios has grown,
the need for transmission protocols that can provide ultra reliable communication while maintaining low latency has become apparent. 
%; see, e.g. \cite{fettweis2014tactile} and \cite{weiner2014design}. 
While it is obvious that increasing the number of relays and/or antennas at each link potentially enables  to attain higher diversity and array gain, utilizing these while meeting stringent latency constraints introduces substantial challenges, one of which is the need for acquiring channel state information rapidly.

%Interestingly, there are scenarios where acquiring transmitter-side CSI is easier than acquiring receiver-side CSI; the transmission phase from the relays to the destination node in the considered setting is among these.
%Namely, 
A reasonable approach is to have the relays acquire local CSI (of both source-to-relay and relay-to-destination links) via channel reciprocity, employing
time-division duplex (TDD). See further discussion in \cite{bletsas2006simple}. %\cite{bletsas2007cooperative}.
Such an approach fits well a scenario where there is a large number of ``potential" relays, but only a rather small subset will be active in a given communication round. Thus, it would be highly inefficient for the source and destination nodes to try to acquire CSI
and then feed the CSI back to the relays.
%, due to the large pool of potential relays.
% This advocates employing transmission strategies that only require  the relays to have  access to local CSI in order to meet strict latency constraints.
We also note that similar considerations lead
to TDD being advocated for use in massive (non-distributed) MIMO systems; see, e.g., \cite{hoydis2013making}.

Accordingly, the proposed technique does not require the transmitter nor the destination node to have access to any CSI. Rather, we assume that perfect, yet local only, CSI is available at each relay.
%, a scenario studied and discussed in depth in e.g.,  \cite{bletsas2006simple}.
We show that 
%given local CSI at the relays, 
provided that the destination node applies the universal space-time diversity combining transformation recently introduced in \cite{domanovitz2018diversity}, then a unitary MIMO channel is induced between each relay and the destination node. 
Each relay can thus invert the channel with no power loss. We demonstrate that this allows to attain outage probabilities that are comparable to those attained when the relays perform centralized beamforming on the maximum singular vector of the joint channel, up to a moderate power loss. 
%Since the resulting channel from each relay to the destination is {\em unitary}, 
The channel inversion operation at the relays can be seen as the analogue of undoing the phase, as performed in distributed beamforming to a single-antenna receiver.

\section{System model}
\label{sec:system}
We assume transmission between a source node and  a destination node  via an array of $M$ relays where the source is equipped with a single antenna and the receiver has $N_r$ antennas.
%as depicted in Figure~\ref{fig:ScenarioNrAnd}.
Our focus will be on the case of $N_r=2$. As for the relays, they may be equipped with an arbitrary number of antennas but for sake of exposition, we will begin by assuming that each is equipped with a single antenna. The extension of the scheme to a scenario where the relays are equipped with multiple antennas is described in Section~\ref{sec:multiple_antenna_relays}.

% For the most part, we will consider the case of $N_r=2$. A discussion of extensions is carried out in Section~\ref{sec:extHighNumOfAnt}.
% The extension of the proposed scheme to relays equipped with multiple antennas is rather straightforward and is addressed in Section~\ref{sec:extHighNumOfAnt}.

We largely assume the system setup as described in \cite{bletsas2006simple} which we therefore only briefly recall, highlighting mostly the difference in assumptions, and referring the reader to the latter works  for more details on the general problem formulation.
The main differences in the assumptions is that in the present work, perfect synchronization of all nodes is assumed and we strive to achieve array gain in addition to diversity gain.
% \begin{itemize}
%     \item In the present work, perfect synchronization of all nodes is assumed and we strive to achieve array gain in addition to diversity gain.
% \item The number of transmit antennas per node is not limited to one and in particular the destination node is equipped with two antennas (or possibly more).
% \end{itemize}
% \begin{remark}
% For simplicity of exposition, we will describe the scheme assuming the destination node has exactly two antennas and source and relay nodes are equipped with one antenna. We note that all the schemes described in the next section can seamlessly be combined with relay transmit antenna selection so as to enjoy further diversity gains (on the forward link) in the case of multiple-antenna relays. As for the link from the source node to relays, simple MRC combining can be used in such a case.
% \end{remark}
% \begin{remark}
% For simplicity of exposition, we will describe the scheme assuming the destination node has exactly two antennas and source and relay nodes are equipped with one antenna.
% \end{remark}

\begin{figure}[htbp]
\begin{center}
    \begin{psfrags}
        \psfrag{A}[][][1]{$\vdots$}
        \psfrag{h11}[][][1]{$h_{11}$}
        \psfrag{h21}[][][1]{$h_{21}$}
        \psfrag{hNr1}[][][1]{$h_{N_r1}$}
        \psfrag{h12}[][][1]{$h_{12}$}
        \psfrag{h22}[][][1]{$h_{22}$}
        \psfrag{hNr2}[][][1]{$h_{N_r2}$}
        \psfrag{h1N}[][][1]{$h_{1M}$}
        \psfrag{h2N}[][][1]{$h_{2M}$}
        \psfrag{hNrN}[][][1]{$h_{N_rM}$}
        \psfrag{g1}[][][1]{${\bf g}_1$}
        \psfrag{g2}[][][1]{${\bf g}_2$}
        \psfrag{g3}[][][1]{${\bf g}_M$}
        \includegraphics[width=0.45\columnwidth]{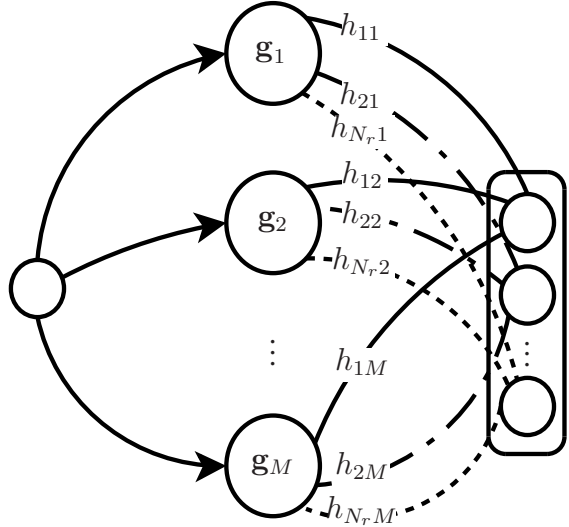}
    \end{psfrags}
\end{center}
\caption{Basic communication scenario between a source node and a destination node via a group of relays.}
\label{fig:ScenarioNrAnd}
\end{figure}

We consider a two-phase protocol.
In the first phase, the source node transmits the coded message and all relays are in listening mode.
As for the second phase, the proposed scheme can equally fit a decode-and-forward (DaF) or
an amplify-and-forward (AaF) mode of operation.
For simplicity, we will describe a DaF protocol
where all nodes that have successfully decoded the message participate in performing distributed beamforming as described next.
We denote the (random) number of relays that successfully decode the message by $M'$.
%We note that in the terminology of \cite{bletsas2007cooperative}, we are considering what is referred to as ``reactive multiple-relay DaF".

We assume that the channel coefficients do not change over the entire transmission period of $2T$ symbols, where each phase occupies $T$ symbols. All nodes are assumed to operate in half-duplex mode and for simplicity we assume there is no direct link between the source and destination.

As for CSI, we assume that before transmission begins, both source node and destination node send a beacon (clear-to-send, ready-to-receive) signal, from which the relays  obtain  \emph{local} CSI (which is assumed to be perfect) by invoking channel reciprocity. Thus, we assume that transmission during both phases takes place over the same frequency band, i.e., we assume TDD.

%To simplify the analysis we assume that all the %relays decode correctly the message sent from %the source (i.e., decode and forward %scheme)\footnote{We note that the scheme we %present next has similar benefits also for %amplify and forward case}.

The source node encodes the data to form the transmitted signal $x(t)$, $t=1,2,\ldots,T$, where $T$ is the blocklength. The transmitted signal must satisfy the power constraint
$\mathbb{E}\{|x(t)|^2\}\leq P_s$.
The received signal at relay $j$ is
\begin{align}
    r_j(t)=h^{s\rightarrow r}_j x(t)+n_j(t),
\end{align}
where $n_j(t)$ is 
%circularly-symmetric complex normal 
$\mathcal{CN}(0,1)$ and
is i.i.d. over time and between relays.
The channel coefficients are distributed in the same manner. Therefore, we may define the nominal SNR between the source and a relay node by
$
    \SNR^{s\rightarrow r} \triangleq P_s.
$

Now each link from a relay to the destination
is a single-input multiple-output (SIMO)
channel with coefficients
\begin{align}
    {\bf h}_j \triangleq {\bf h}^{r \rightarrow d}_j \triangleq \begin{bmatrix} h_{1j} & h_{2j} & \cdots & h_{N_rj}\end{bmatrix}^T,
\end{align}
for $j=1,\ldots,M'$.
%, where $M'$ denotes the number of relays that have successfully decoded the message.
Without loss of generality, we  assume that the relays with indices $1,\ldots,M'$ are the ``successful" relays.

We denote the symbols sent from the relays by $x_j(t)$ where 
%and assume that each active relay must satisfy the (individual) 
each relay is subject to the power constraint  $\mathbb{E}\{|x_j|^2\}=P_{r,i}$.
For simplicity, we further assume that
$P_{r,i}=P_r$ for all $j$.
Thus, signal received at the destination is given by
\begin{align}
    {\bf s}_{\rm dest}(t)=\sum_{j=1}^{M'} {\bf h}_j x_j(t)+{\bf n}(t),
    \label{sdest}
\end{align}
for $t=1,\ldots,T$, and where  $n_j(t)$ is i.i.d. $\mathcal{CN}(0,1)$ (over space and time).
We note that with a slight abuse of notation we let $t$ run form $1$ to $T$ in both phases of transmission.
We define the nominal SNR between a relay node and the destination by
$
    \SNR^{r\rightarrow d} \triangleq P_r.
$

We now describe the second phase of transmission. Each of the $M'$ relays that have successfully decoded the message has access to the transmitted symbols $x(t)$, $t=1,\ldots,T$. We will only consider relaying operations that amount to applying a linear transformation to the received codeword.  We do assume that buffering of symbols is possible  and hence linear space-time modulation can be applied at the relay.

Nonetheless, for simplicity, we first describe the simplest setting (without buffering), in which case the operation done at each relay amounts to multiplying each codeword symbol by some complex number which we take to be independent of $t$. We denote this scalar by $g_j$, $j=1,\ldots,M'$.
Thus, the output of each relay is simply
$$x_j(t)=g_j x(t), \quad t=1,\ldots,T,$$
and hence the destination node receives
\begin{align}
\svv{s}(t)=\sum_{j=1}^{M'} g_j{\bf h}_j   x(t)+\svv{n}(t).
\label{eq:rec_sig}
\end{align}
% We relays has perfect local channel state information (CSI) which can be obtained via reciprocity. The focus of this paper is with distributed processing hence we assume no information exchange between relays is possible. Yet, we analyze the performance achieved with centralized processing as a benchmark for the performance of the distributed schemes.
% We assume that the signals transmitted from the source are subject to the power constraint $\mathbb{E}\{|x|^2\}=P$. We further assume that
% the noise
%${\bf n}(t)$  is i.i.d. %$\mathcal{CN}(0,1)$.
%over space and time. %with
% samples that are circularly-symmetric complex Gaussian.
Defining
\begin{align}
    \alpha \triangleq\sqrt{P_r/P_s},
    \label{def:alpha}
\end{align}
it follows that the gains $g_j$ should be chosen such that $|g_j|=\alpha$.
When considering
more general space-time processing at the relays, \eqref{eq:rec_sig} is replaced with a corresponding matrix variant as described in the sequel.
% \begin{remark}
% In case of multiple-antenna relays, since we assume local CSI, for the link from the source node to relays, simple MRC combining can be used. For the forward link, all the schemes described in the next section can seamlessly be combined with beamforming applied to the channel vector from the relay to the destination to enjoy further diversity and array gains.

% [THE REST SHOULD BE REMARK TWO]
% Moreover,  destination has more than a single receive antenna, each relay should transmit at the direction of the singular value which corresponds the strongest singular value.
% % We note that all the schemes described in the next section can seamlessly be combined with relay transmit antenna selection so as to enjoy further diversity gains (on the forward link) in the case of multiple-antenna relays. As for the link from the source node to relays, simple MRC combining can be used in such a case.
% \end{remark}

% Second is denoted ``long-trem'' power constraint in which the average power transmitted by {\emph all relays} is subject to the power constraint
% \begin{align}
%     \sum_{i=1}^{N}E\{|x_j|^2\} = NP.
% \end{align}
% This suggests the instantaneously, a relay can utilize a transmit power which is much larger than the average.

We will compare the outage probability attained by different schemes and take as a figure of merit, the receive SNR attained at the destination node.
This can be directly translated to an outage probability for either uncoded transmission or coded transmission, depending on the stringency of the latency constraints.
%Both cases will be analyzed.
% different schemes. We further interested in low latency schemes. To demonstrate the performance of different schemes under the low latency constraint we demonstrate their performance while assuming uncoded transmission.
In particular, in order to provide simple performance bounds, we will analyze the mutual information attained by a  scheme, defined by
\begin{align}
    I(\SNR_{\rm scheme}) \triangleq\log(1+\SNR_{\rm scheme}).
\end{align}
Correspondingly,  for coded transmission (with long blocklength), outage is defined as the event where the mutual information is below the target rate $R_{\rm tar}$, i.e.
\begin{align}
    \prob\left(I(\SNR_{\rm scheme})<R_{\rm tar}\right).
\end{align}

\section{New distributed beamforming protocol}
\label{sec:protocol}
The proposed method utilizes a recently introduced
diversity combining transformation \cite{domanovitz2018diversity} that is performed as a front-end operation at the destination node.  This transformation, briefly recalled next, may be viewed as the dual of Alamouti modulation \cite{alamouti1998simple}.
%developed universal diversity combining scheme that we employ as a front-end operation at the destination node. %We therefore begin by briefly recalling the scheme presented in \cite{domanovitz2018diversity}.
%[UPDATE CITATION WITH PUBLICATION DATE IN TCOM]

\subsection{Universal Diversity Combining Transformation 
%(Single User)
}
\label{sec:transformation}
Consider a $2 \times 1$ single-input multiple-output (SIMO) channel, with channel coefficients $h_1$ and $h_2$. The signal received at antenna $j=1,2$, at
discrete time $t$, is
\begin{align}
    s_{j}(t)=h_j x(t)+n_j(t).
    \label{eq:receive_sigs}
\end{align}
We assume that
the noise $n_j(t)$ is $\mathcal{CN}(0,1)$ and
is i.i.d. over space and time.
%samples that are circularly-symmetric %complex Gaussian random variables with %unit variance. 
We further assume the transmitted symbols are subject to the power constraint $\mathbb{E}\{|x|^2\}=P$.
%\footnote{When transmission of multiple users is considered, we assume equal-power isotropic transmission, i.e., $\mathbb{E}(\bf{x}^T\bf{x})=\svv{I}\cdot P$.}

The scheme works on batches of two time instances and
for our purposes, it will suffice to describe it for
time instances $t=1,2$. Let us stack the four complex samples received over $T=2$ time instances, two over each antenna, into an $8 \times 1$
real vector:
\begin{align*}
\mathbf{s}=[s_{1R}(1)  s_{1I}(1) s_{2R}(1) s_{2I}(1) s_{1R}(2)   s_{1I}(2)   s_{2R}(2)  s_{2I}(2)   ]^T,
%\label{eq:s_vector}
\end{align*}
where $x_R$ and $x_I$ denote the real and imaginary parts of a complex number $x$.
We similarly define the stacked noise vector $\svv{n}$.
Likewise, we define
\begin{align}
\mathbf{x}=[x_{R}(1) \, x_{I}(1)  \, x_{R}(2) \,  x_{I}(2)]^T.
\label{eq:source_vec}
\end{align}

Next, we form a real vector with $4$ elements $\svv{r}$ by applying to the vector $\mathbf{s}$ the transformation $\mathbf{r}=\svv{\PrecMat} \mathbf{s}$
% \begin{align}
% \mathbf{y}=\svv{\PrecMat} \mathbf{s}
% \label{eq:rec_trans}
% \end{align}
where
\begin{align}
\svv{\PrecMat}=\frac{1}{\sqrt{2}}
  \left[ {\begin{array}{cccccccc}
  1 & 0 & 0 & 0 &  0 & 0 &  1 &  0 \\
  0 & 1 & 0 & 0 &  0 & 0 &  0 & -1 \\
  0 & 0 & 1 & 0 & -1 & 0 &  0 &  0 \\
  0 & 0 & 0 & 1 &  0 & 1 &  0 &  0
  \end{array} } \right].
  \label{eq:P}
\end{align}
Note that unlike conventional diversity-combining schemes, here the combining matrix
$\svv{\PrecMat}$ is \emph{universal}, i.e., it does not depend on the channel coefficients.

% As shown in the Appendix, that the following holds.
In \cite{domanovitz2018diversity} it is shown that the following relation holds,
\begin{align}
    %\svv{y}&=\frac{\|\svv{h}\|}{\sqrt{2}}\svv{O}(h_1,h_2)\svv{x}+\svv{\PrecMat} \svv{n}  \nonumber \\
    \svv{r}=\frac{\|\svv{h}\|}{\sqrt{2}}\svv{O}(h_1,h_2)\svv{x}+\svv{n'},
    \label{eq:n'}
\end{align}
where
\begin{align}
\svv{O}(h_1,h_2)=\frac{1}{\|\svv{h}\|}
  \left[ {\begin{array}{cccc}
  h_{1R} & -h_{1I}  &  h_{2R} & -h_{2I}  \\
  h_{1I} &  h_{1R}  & -h_{2I} & -h_{2R}  \\
  h_{2R} & -h_{2I}  & -h_{1R} &  h_{1I}  \\
  h_{2I} &  h_{2R}  &  h_{1I} &  h_{R1}
  \end{array} } \right]
  \label{eq:Ueff}
\end{align}
is an orthonormal matrix for any realization of $h_1,h_2$, and where $\svv{n'}$ is i.i.d.
and Gaussian with variance $1/2$.\footnote{The variance is
$1/2$ as we chose above to normalize the complex noise to have unit power.}

\subsection{Proposed Distributed Beamforming Scheme}
% In the proposed scheme, the relays 
% %a scalar source which is processed at $M'$ relays prior to transmission to the destination.
% the destination node applies the universal space-time diversity transformation as a front end-operation. Hence, \eqref{eq:rec_vector} becomes
% \begin{align}
% \svv{y}&=\sum_{j=1}^{M'}\frac{\|\svv{h}_j\|}{\sqrt{2}}\svv{O}(h_{1j},h_{2j})\svv{x}_j+\svv{n}'.
% \end{align}

We consider now the scenario of a $2\times K$ MIMO multiple-access channel (MIMO-MAC) where $M'$ users, each equipped with a single antenna, transmit to a common receiver that is equipped with two antennas. The input/output relation of this MIMO-MAC is given in \eqref{sdest}.

%can be expressed as
%\begin{align}
%    \svv{s}(t)=\sum_{j=1}^{M'}\svv{h}_j %x_j(t)+\svv{n}(t),
%    \label{eq:MIMO_MAC}
%\end{align}
%where $\svv{h}_j$ is the $2\times 1$ channel matrix  between user $j$ and the receiver, 
%We assume isotropic (``white'') transmission by each user and that all users are subject to the 
%all users subject to the same power constraint $P$. The noise $\svv{n}(t)$ is $\mathcal{CN}(0,1)$ and
%is i.i.d. over space and time.
%with
%samples that are circularly-symmetric complex %Gaussian random variables with unit variance.

Now assume that the receiver applies as a front end the universal diversity combining transformation described in Section~\ref{sec:transformation}, applied over two consecutive time instances. Then by \eqref{sdest} and (\ref{eq:n'}), in a MAC scenario, the receiver output is 
\begin{align}
\svv{y}&=\sum_{j=1}^{M'}\frac{\|\svv{h}_j\|}{\sqrt{2}}\svv{O}(h_{1j},h_{2j})\svv{x}_j+\svv{n}',
\label{eq:rec_vector}
\end{align}
where $\svv{h}_k=\begin{bmatrix}h_{1k} & h_{2k}\end{bmatrix}^T$ and $\svv{O}(h_{1j},h_{2j})$ is given by \eqref{eq:Ueff}.

Since we assume that each relay has perfect local CSI, at the expense of adding an additional delay of one symbol, in the proposed scheme, each relay that decoded the message successfully, simply ``undoes" its channel matrix and forwards the received (noiseless) symbols. Specifically, each relay transmits
\begin{align}
    {\bf x}_j & = \alpha\svv{O}(h_{1j},h_{2j})^{-1} {\bf x} \nonumber \\
    & = \alpha\svv{O}(h_{1j},h_{2j})^T {\bf x}
    \label{eq:invAtRelay}
\end{align}
where ${\bf x}$ is defined in \eqref{eq:source_vec} and should be interpreted as two consecutive symbols transmitted from the source node,
%and correctly decoded at the $M'$ relays %participating in transmission,  
$\svv{O})(\cdot,\cdot)$ is defined in \eqref{eq:Ueff}, and $\alpha$ is defined in \eqref{def:alpha}.
%Note that the per-antenna power constraint is satisfied due to the scaling factor $\alpha$ which is defined in \eqref{def:alpha}.

Thus, the destination sees the effective channel
\begin{align}
   % \svv{y}&=\sum_{j=1}^{M'}\alpha\frac{\|\svv{h}_j\|}{\sqrt{2}}\svv{O}(h_{1j},h_{2j}){\bf x}_j+\svv{n}' \nonumber \\
   \svv{y}&=\sum_{j=1}^{M'}\alpha\frac{\|\svv{h}_j\|}{\sqrt{2}}\svv{O}(h_{1j},h_{2j})\svv{O}(h_{1j},h_{2j})^T {\bf x}+\svv{n}' \nonumber \\
    &=\sum_{j=1}^{M'}\alpha\frac{\|\svv{h}_j\|}{\sqrt{2}} {\bf x}+\svv{n}'.
\end{align}
The latter is a set of independent scalar channels with
%that can be decoded \emph{without any CSI at the receiver}.  SNR is
\begin{align}
    \SNR=\displaystyle{\frac{\left(\sum_{j=1}^{M'}\|\svv{h}_j\|\right)^2}{2}P_r}.
    \label{eq:finalSNROfNewMethod}
\end{align}

The attained SNR is quite pleasing as we obtain both the maximal diversity gain while also enjoying transmit-side array gain (but not receive-side MRC gain).

Equivalently, using the SVD of the channel between the $j$'th relay and the destination
$\svv{h}_j=\svv{U}_j\begin{bmatrix} d_j \\ 0 \end{bmatrix}v_j^*$,
we may rewrite \eqref{eq:finalSNROfNewMethod} as
% Therefore, \eqref{eq:finalSNROfNewMethod} can be written as
$\SNR=\displaystyle{\frac{\left(\sum_{j=1}^{M'}d_j\right)^2}{2}P_r}$.

\subsection{Extension to Multi-Antenna Relays}
\label{sec:multiple_antenna_relays}
% In case the relays or destination has more than a single antenna (we still assume that the source is equipped with a single antenna and thus only a single stream is transmitted), the suggested method can be extended by first converting each MIMO channel from a relay to the destination is converted to a SIMO channel and then use previously suggested methods for performing dimension reduction transformation to the case of a receiver with more than two antennas.

% As detailed in \cite{domanovitz2018diversity}, all of the dimension reduction transformations (when $N_r>2$) result in a non-orthogonal effective channel and thus suffer from some additional loss in performance with respect to MRC or antenna selection.  Nevertheless, employing such transformations is still valuable as no CSI is needed.

% Assuming the destination node has $N_r$ antennas (while each relay has one transmit antenna), and denoting by $\mathcal{F}$ the effective channel resulting from applying the universal transformation,  the received signal is
% \begin{align}
% \svv{y}&=\sum_{j=1}^{M'}\frac{\|\svv{h}_j\|}{c}\mathcal{F}(h_{1j}\cdots,h_{N_r j})\svv{x}_j+\svv{n}',
% \end{align}
% where $c$ is a power normalization constant that depends (only) on the chosen transformation. See \cite{domanovitz2018diversity} for a detailed example for the case of $4$ receive antennas.

When the relays have multiple antennas, we note that any beamforming vector (meeting the per-relay power constraint) can be applied at each relay in conjunction with the universal dimension reduction operation at the destination node. 
%We demonstrate this we start with the case where each relay has $N_t$ antennas and the destination has two antennas. 
Assuming for simplicity that the relays are equipped with the same number of antenna, denote by $\svv{H}_j$  the $2\times N_t$ channel from the $j$'th relay to the destination, with SVD 
%The SVD of this channel takes the form
\begin{align}
    \svv{H}_j=\svv{U}_j\begin{bmatrix} d_{i1} & 0 & \cdots & 0 \\ 0 & d_{i2} & \cdots & 0 \end{bmatrix}\svv{V}_j^H
\end{align}
Applying beamforming vector $\bf{p}=\begin{bmatrix} p_1 & \cdots & p_{N_t} \end{bmatrix}$$^T$ (where $\sum|p_j|^2=1$), we get
\begin{align}
    \svv{H}_j\bf{p}&=\svv{U}_j\begin{bmatrix} d_{j1} & 0 & \cdots & 0 \\ 0 & d_{j2} & \cdots & 0 \end{bmatrix}\svv{V}_j^H\begin{bmatrix} p_1 & \cdots & p_{N_t} \end{bmatrix}^T \nonumber \\
    &= \svv{U}_j\begin{bmatrix} d_{j1} & 0 & \cdots & 0 \\ 0 & d_{j2} & \cdots & 0 \end{bmatrix}\begin{bmatrix} \tilde{p}_1 & \cdots & \tilde{p}_{N_t} \end{bmatrix}^T \nonumber \\
    & = \svv{U}_j\begin{bmatrix} d_{j1}\tilde{p}_1 \\ d_{j2}\tilde{p}_2 \end{bmatrix}
    % &= \svv{U}_i
\end{align}
where $|\tilde{p}_1|^2+|\tilde{p}_2|^2 \leq 1$ (the result of a unitary transformation on the unit norm  vector $\bf{p}$). The application of the  beamforming vector transforms the $2\times N_t$ MIMO channel to a $2\times 1$ SIMO channel to which the universal dimension reduction transformation can be applied. Furthermore, it is readily seen that the singular value of this MISO channel is $\tilde{d}_j=\sqrt{(d_{j1}\tilde{p}_1)^2+(d_{j2}\tilde{p}_2)^2}$.
It is easily shown that the optimal beamforming vector is the right singular vector of $\svv{H}_j$ corresponding to the largest singular value.

\section{Numerical Performance Evaluation}
\label{sec:performance}
%We now numerically compare the performance of different schemes when operating in an i.i.d.  Rayleigh fading environment. 

%As mentioned above, we consider the outage probability of the mutual information.

%as well as the outage probability for uncoded transmission. %Figure~\ref{fig:miFig_M4_R4_SNR_1ant} depicts the outage probability of the mutual information for different topologies and schemes

\begin{figure}[htbp]
    \centering
    \includegraphics[width=\figSize\columnwidth]{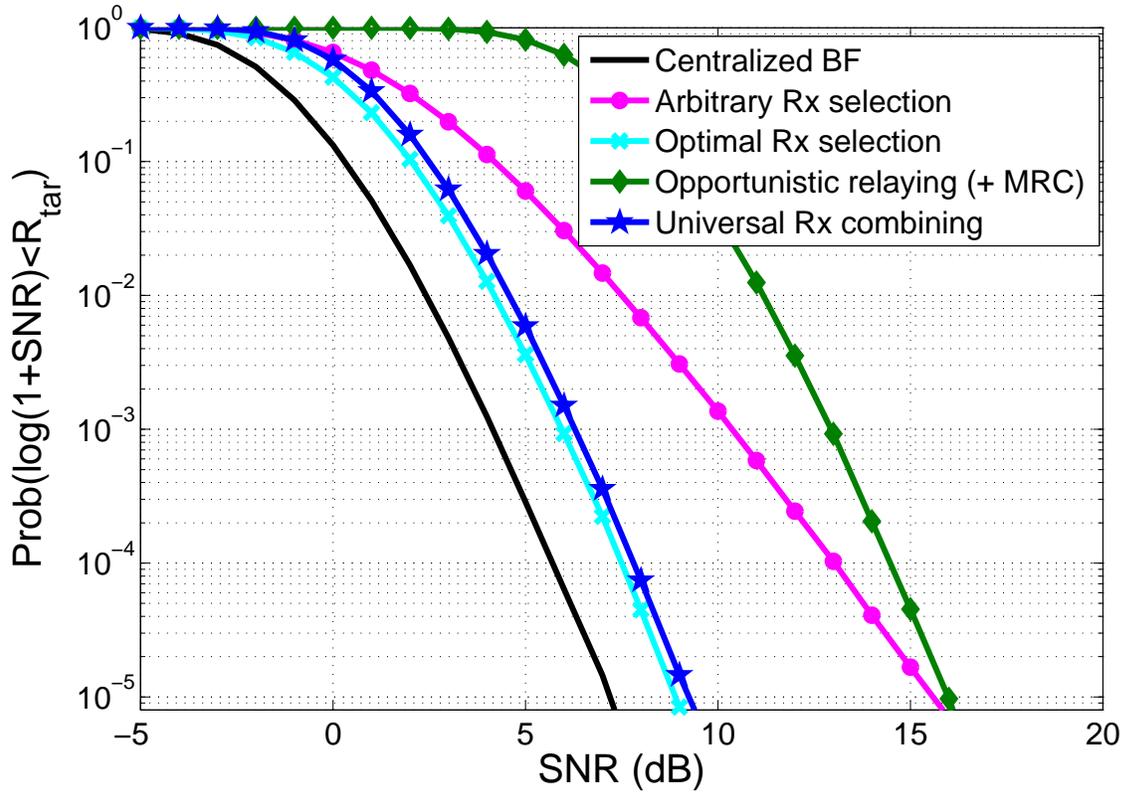}
    \caption{Outage probability of mutual information for i.i.d. Rayleigh fading with $M'=4$ relays, $R_{\rm tar}=4$, as a function of the transmit power at the relays $P_r$.
    %for a  receiver equipped with two antennas.
    }
    \label{fig:OutageR4_4relays_SNR_twoAntRec}
\end{figure}

%\subsection{Evaluation of performance over second transmission phase}
We compare the performance of the proposed method with several different schemes considering an i.i.d. Rayleigh fading environment where each of the relays has a single transmit antenna.
%and the receiver has two antennas. 
We first consider a scenario in which exactly $M'=4$ relays participate in the second phase of transmission.
Figure~\ref{fig:OutageR4_4relays_SNR_twoAntRec} depicts the outage probability of the mutual information.  As simple benchmarks that are compatible with low latency constraints, we consider arbitrary antenna selection at the destination, which results in
    $$\SNR=\left(\sum_{j=1}^{M'} |h_{1j}| \right)^2 P_r$$
    and opportunistic relaying, which results in\footnote{Note that for the latter, in order to benefit from the MRC gain of the two antennas at the receiver, additional training required to estimate the channel, once the best relay is chosen.}
    $$\SNR=\max_{j\in\{1,\ldots,M'\}}\sum_{i=1}^{2} |h_{ij}|^2 P_r.$$
Both methods suffer a significant penalty in terms of the transmit power required to meet a given  outage probability as compared with  centralized beamforming. As a further benchmark (which may not meet the low latency constraint) we take optimal (receiver) antenna selection which results in
$$\SNR=\max_{i\in\{1,2\}}\left(\sum_{j=1}^{M'} |h_{ij}| \right)^2 P_r.$$
As can be seen, the proposed method suffers only a small loss with respect to optimal (receiver) antenna selection.

% Figure~\ref{fig:OutageR4_4relays_SNR_twoAntRec} also depicts methods \emph{that do not meet} the low latency requirement or per-relay power constraint. Namely, optimal selection (which does not meet the former requirement) and optimal opportunistic relaying subject only to a sum power constraint  (which does not meet the latter requirement) are depicted.
% These result in a much smaller gap with respect to centralized beamforming.
% It can be seen that the proposed method achieves similar performance while meeting both these requirements. The performance of ideal distributed space-time coding is also depicted in Figure~\ref{fig:OutageR4_4relays_SNR_twoAntRec}. Even without taking into account additional penalties, the performance of this scheme is significantly worse than the proposed method (and the gap will be larger if compared against differential distributed space-time coding).  
%Figure~\ref{fig:SER_4relays_SNR_twoAntRec} demonstrates a similar behavior for uncoded QPSK transmission.

\begin{figure}[htbp]
    \centering
    \includegraphics[width=\figSize\columnwidth]{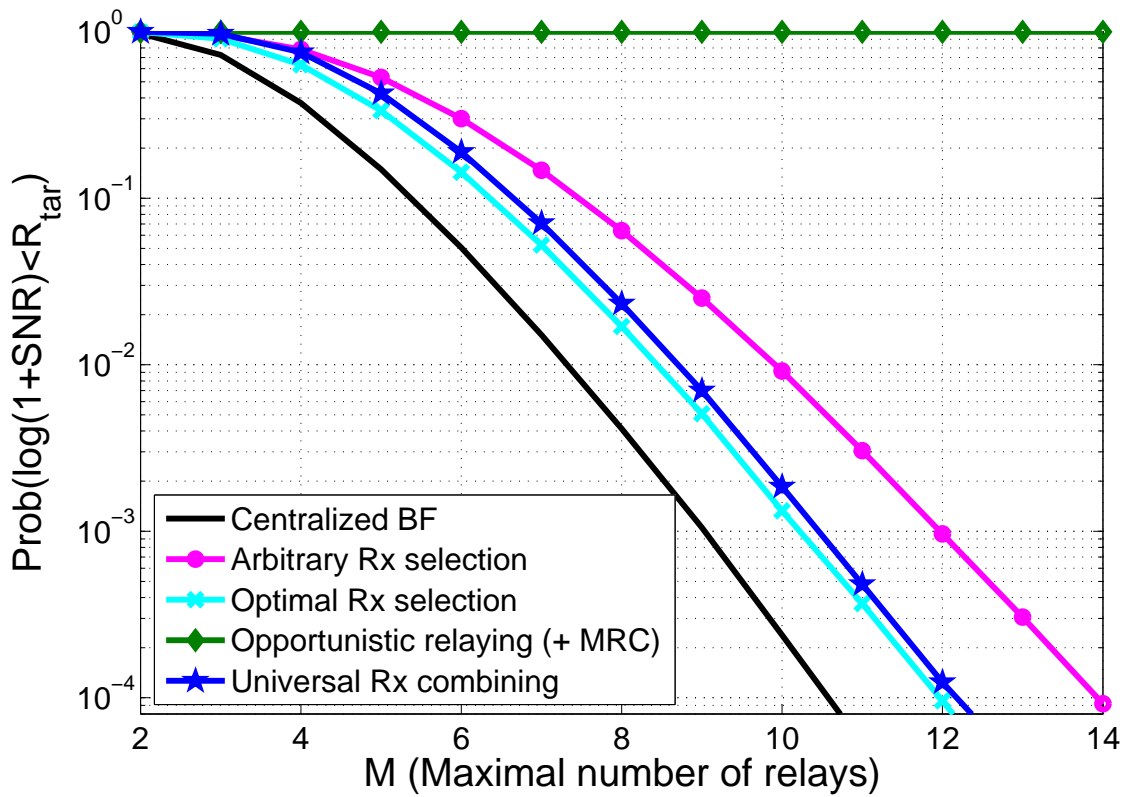}
    \caption{Outage probability of mutual information for i.i.d. Rayleigh fading  with $R_{\rm tar}=4$, $P_s=20$ dB,  $P_r=0$ dB, as a function of the total number of relays (the number $M'$ of  active ones being random). 
    %for a  receiver equipped with two antennas.
    }
    \label{fig:MI_SNR1_20_SNR2_0}
\end{figure}

We next simulated the end-to-end performance when both phases of transmission are in operation. That is, we now include the first hop in the simulation.
Figures~\ref{fig:MI_SNR1_20_SNR2_0} show the performance of the different schemes as a function of the  maximal possible number of relays $M$, where we set $P_s=20$ dB and $P_r=0$ dB. In these figures, the number of active relays $M'$ is a random variable that depends on the SNR of the links between the source node and the relays. It is worth noting that increasing the (effective) number of active relays boosts the performance of schemes that enjoy array gain in the second transmission phase.
%has an immediate impact on the performance of schemes which perform selection or combining at the receiver side (since all enjoy higher array gain). 
In contrast, opportunistic relaying, which does not exploit transmit-side CSI in the second phase, can only gain from enhanced diversity when the number of relays grows, resulting in very poor performance.
%are active hence the improvement in its performance is much lower as the number of relays increases.
% Again, outage probability of the mutual information in depicted in Figure \ref{fig:MI_4relays_SNR} while the symbol error rate of uncoded transmission is depicted in Figure \ref{fig:uncoded_4relays_SNR}.

\bibliographystyle{IEEEtran}
\bibliography{eladd}

\begin{thebibliography}{10}
\providecommand{\url}[1]{#1}
\csname url@samestyle\endcsname
\providecommand{\newblock}{\relax}
\providecommand{\bibinfo}[2]{#2}
\providecommand{\BIBentrySTDinterwordspacing}{\spaceskip=0pt\relax}
\providecommand{\BIBentryALTinterwordstretchfactor}{4}
\providecommand{\BIBentryALTinterwordspacing}{\spaceskip=\fontdimen2\font plus
\BIBentryALTinterwordstretchfactor\fontdimen3\font minus
  \fontdimen4\font\relax}
\providecommand{\BIBforeignlanguage}[2]{{%
\expandafter\ifx\csname l@#1\endcsname\relax
\typeout{** WARNING: IEEEtran.bst: No hyphenation pattern has been}%
\typeout{** loaded for the language `#1'. Using the pattern for}%
\typeout{** the default language instead.}%
\else
\language=\csname l@#1\endcsname
\fi
#2}}
\providecommand{\BIBdecl}{\relax}
\BIBdecl

\bibitem{sendonaris2003user1}
A.~Sendonaris, E.~Erkip, and B.~Aazhang, ``User cooperation diversity. {P}art
  {I}. {S}ystem description,'' \emph{IEEE Transactions on Communications},
  vol.~51, no.~11, pp. 1927--1938, Nov 2003.

\bibitem{sendonaris2003user2}
------, ``User cooperation diversity. {P}art {II}. {I}mplementation aspects and
  performance analysis,'' \emph{IEEE Transactions on Communications}, vol.~51,
  no.~11, pp. 1939--1948, Nov 2003.

\bibitem{laneman2004cooperative}
J.~N. Laneman, D.~N.~C. Tse, and G.~W. Wornell, ``Cooperative diversity in
  wireless networks: Efficient protocols and outage behavior,'' \emph{IEEE
  Transactions on Information Theory}, vol.~50, no.~12, pp. 3062--3080, Dec
  2004.

\bibitem{laneman2003distributed}
J.~N. Laneman and G.~W. Wornell, ``Distributed space-time-coded protocols for
  exploiting cooperative diversity in wireless networks,'' \emph{IEEE
  Transactions on Information Theory}, vol.~49, no.~10, pp. 2415--2425, Oct
  2003.

\bibitem{bletsas2006simple}
A.~Bletsas, A.~Khisti, D.~P. Reed, and A.~Lippman, ``A simple cooperative
  diversity method based on network path selection,'' \emph{IEEE {J}ournal on
  {S}elected {A}reas in {C}ommunications}, vol.~24, no.~3, pp. 659--672, March
  2006.

\bibitem{bletsas2007cooperative}
A.~Bletsas, H.~Shin, and M.~Z. Win, ``Cooperative communications with
  outage-optimal opportunistic relaying,'' \emph{IEEE Transactions on Wireless
  Communications}, vol.~6, no.~9, pp. 3450--3460, September 2007.

\bibitem{hunter2002cooperation}
T.~E. Hunter and A.~Nosratinia, ``Cooperation diversity through coding,'' in
  \emph{Proceedings IEEE International Symposium on Information Theory,}, 2002,
  p. 220.

\bibitem{stefanov2002cooperative}
A.~Stefanov and E.~Erkip, ``Cooperative information transmission in wireless
  networks,'' in \emph{Proceedings of the IEEE Asian-European Information
  Theory Workshop, Breisach, Germany}, June, 2002.

\bibitem{kramer2005cooperative}
G.~Kramer, M.~Gastpar, and P.~Gupta, ``Cooperative strategies and capacity
  theorems for relay networks,'' \emph{IEEE Transactions on Information
  Theory}, vol.~51, no.~9, pp. 3037--3063, Sept 2005.

\bibitem{scaglione2003opportunistic}
A.~Scaglione and Y.-W. Hong, ``Opportunistic large arrays: Cooperative
  transmission in wireless multihop ad hoc networks to reach far distances,''
  \emph{IEEE transactions on Signal Processing}, vol.~51, no.~8, pp.
  2082--2092, 2003.

\bibitem{nabar2004fading}
R.~U. Nabar, H.~Bolcskei, and F.~W. Kneubuhler, ``Fading relay channels:
  Performance limits and space-time signal design,'' \emph{IEEE Journal on
  Selected Areas in communications}, vol.~22, no.~6, pp. 1099--1109, 2004.

\bibitem{fettweis2014tactile}
G.~P. Fettweis, ``The tactile internet: Applications and challenges,''
  \emph{IEEE Vehicular Technology Magazine}, vol.~9, no.~1, pp. 64--70, 2014.

\bibitem{weiner2014design}
M.~Weiner, M.~Jorgovanovic, A.~Sahai, and B.~Nikoli{\'e}, ``Design of a
  low-latency, high-reliability wireless communication system for control
  applications,'' in \emph{Communications (ICC), 2014 IEEE International
  Conference on}, 2014, pp. 3829--3835.

\bibitem{jing2006distributed}
Y.~Jing and B.~Hassibi, ``Distributed space-time coding in wireless relay
  networks,'' \emph{IEEE Transactions on Wireless Communications}, vol.~5,
  no.~12, pp. 3524--3536, 2006.

\bibitem{jing2007using}
Y.~Jing and H.~Jafarkhani, ``Using orthogonal and quasi-orthogonal designs in
  wireless relay networks,'' \emph{IEEE Transactions on Information Theory},
  vol.~53, no.~11, pp. 4106--4118, 2007.

\bibitem{jing2008distributed}
------, ``Distributed differential space-time coding for wireless relay
  networks.'' \emph{IEEE Trans. Communications}, vol.~56, no.~7, pp.
  1092--1100, 2008.

\bibitem{larsson2014massive}
E.~G. Larsson, O.~Edfors, F.~Tufvesson, and T.~L. Marzetta, ``Massive {MIMO}
  for next generation wireless systems,'' \emph{IEEE Communications Magazine},
  vol.~52, no.~2, pp. 186--195, 2014.

\bibitem{hoydis2013making}
J.~Hoydis, K.~Hosseini, S.~t. Brink, and M.~Debbah, ``Making smart use of
  excess antennas: Massive {MIMO}, small cells, and {TDD},'' \emph{Bell Labs
  Technical Journal}, vol.~18, no.~2, pp. 5--21, 2013.

\bibitem{brennan1959linear}
D.~G. Brennan, ``Linear diversity combining techniques,'' \emph{Proceedings of
  the IRE}, vol.~47, no.~6, pp. 1075--1102, 1959.

\bibitem{zheng2007mimo}
X.~Zheng, Y.~Xie, J.~Li, and P.~Stoica, ``{MIMO} transmit beamforming under
  uniform elemental power constraint,'' \emph{IEEE Transactions on Signal
  Processing}, vol.~55, no.~11, pp. 5395--5406, 2007.

\bibitem{domanovitz2018diversity}
E.~Domanovitz and U.~Erez, ``Diversity combining via universal
  dimension-reducing space-time transformations,'' \emph{IEEE Transactions on
  Communications}, 2018.

\bibitem{alamouti1998simple}
S.~M. Alamouti, ``A simple transmit diversity technique for wireless
  communications,'' \emph{IEEE Journal on Selected Areas in Communications},
  vol.~16, no.~8, pp. 1451--1458, 1998.

\bibitem{pun2009opportunistic}
M.-O. Pun, D.~R. Brown, and H.~V. Poor, ``Opportunistic collaborative
  beamforming with one-bit feedback,'' \emph{IEEE Transactions on Wireless
  Communications}, vol.~8, no.~5, 2009.

\bibitem{tarokh1999space}
V.~Tarokh, H.~Jafarkhani, and A.~R. Calderbank, ``Space-time block codes from
  orthogonal designs,'' \emph{IEEE Transactions on Information Theory},
  vol.~45, no.~5, pp. 1456--1467, 1999.

\bibitem{adams2007minimum}
S.~S. Adams, N.~Karst, and J.~Pollack, ``The minimum decoding delay of maximum
  rate complex orthogonal space--time block codes,'' \emph{IEEE transactions on
  information theory}, vol.~53, no.~8, pp. 2677--2684, 2007.

\bibitem{adams2011novel}
S.~S. Adams, J.~Davis, N.~Karst, M.~K. Murugan, B.~Lee, M.~Crawford, and
  C.~Greeley, ``Novel classes of minimal delay and low {PAPR} rate 1/2 complex
  orthogonal designs,'' \emph{IEEE Transactions on Information Theory},
  vol.~57, no.~4, pp. 2254--2262, 2011.

\bibitem{liu2015minimum}
X.~Liu, Y.~Li, and H.~Kan, ``On the minimum decoding delay of balanced complex
  orthogonal designs,'' \emph{IEEE Transactions on Information Theory},
  vol.~61, no.~1, pp. 696--699, 2015.

\bibitem{tse2005fundamentals}
D.~Tse and P.~Viswanath, \emph{Fundamentals of wireless communication}.\hskip
  1em plus 0.5em minus 0.4em\relax Cambridge university press, 2005.

\bibitem{jafarkhani2001quasi}
H.~Jafarkhani, ``A quasi-orthogonal space-time block code,'' \emph{IEEE
  Transactions on Communications}, vol.~49, no.~1, pp. 1--4, 2001.

\bibitem{tirkkonen2000minimal}
O.~Tirkkonen, A.~Boariu, and A.~Hottinen, ``Minimal non-orthogonality rate 1
  space-time block code for 3+ {T}x antennas,'' in \emph{Spread Spectrum
  Techniques and Applications, 2000 IEEE Sixth International Symposium on},
  vol.~2.\hskip 1em plus 0.5em minus 0.4em\relax IEEE, 2000, pp. 429--432.

\bibitem{sharma2003improved}
N.~Sharma and C.~B. Papadias, ``Improved quasi-orthogonal codes through
  constellation rotation,'' \emph{IEEE Transactions on Communications},
  vol.~51, no.~3, pp. 332--335, 2003.

\end{thebibliography}

\end{document}